\documentclass[aps,twocolumn,preprintnumbers,prl,amsmath,
amssymb,amsfonts,superscriptaddress,floatfix]{revtex4}

\usepackage{amsfonts}
\usepackage{amssymb}
\usepackage{amsmath}
\usepackage{ulem}
\usepackage{color}
\usepackage{latexsym}
\usepackage{graphicx}
\usepackage{graphics}
\usepackage{floatflt}
\usepackage{epsfig}
\usepackage{overpic}
\usepackage{soul} 
\usepackage[latin1]{inputenc}

\newcommand{\be}{\begin{equation}}
\newcommand{\ee}{\end{equation}}
\newcommand{\bea}{\begin{eqnarray}}
\newcommand{\eea}{\end{eqnarray}}

\def\lb{\label}

\def\a{\alpha}
\def\b{\beta}

\newcount\bozza \bozza=0
\ifnum\bozza=1
\newdimen\shift \shift=-2truecm
\def\lb#1{%
{\label{#1}\rlap{\kern\shift{$\scriptstyle#1$}}}}
\else\def\lb#1{\label{#1}} \fi

\begin{document}
\title{Photoinduced nematic state in FeSe$_{0.4}$Te$_{0.6}$} 
\author{Laura Fanfarillo} 
\affiliation{International School for Advanced Studies 
(SISSA) and CNR-IOM, Via Bonomea 265, I-34136, Trieste, Italy}

\author{Damir Kopi\'{c}} 
\affiliation{Dipartimento di Fisica, Università degli Studi di Trieste, 34127 Trieste, Italy}

\author{Andrea Sterzi} 
\affiliation{Dipartimento di Fisica, Università degli Studi di Trieste, 34127 Trieste, Italy}

\author{Giulia Manzoni} 
\affiliation{Dipartimento di Fisica, Università degli Studi di Trieste, 34127 Trieste, Italy}

\author{Alberto Crepaldi} 
\affiliation{Institute of Physics, Ecole Polytechnique Fédérale de Lausanne (EPFL), CH-1015 Lausanne, Switzerland}
\affiliation{Lausanne Centre for Ultrafast Science (LACUS), Ecole Polytechnique Fédérale de Lausanne (EPFL), CH-1015 Lausanne, Switzerland}

\author{Vladimir Tsurkan}
\affiliation{Center for Electronic Correlations and Magnetism, Experimental Physics V, 	University of Augsburg, D-86159 Augsburg, Germany}
\affiliation{Institute of Applied Physics, MD 2028 Chisinau, Republic of Moldova}

\author{Dorina Croitori}
\affiliation{Institute of Applied Physics, MD 2028 Chisinau, Republic of Moldova}

\author{Joachim Deisenhofer} 
\affiliation{Center for Electronic Correlations and Magnetism, Experimental Physics V, 	University of Augsburg, D-86159 Augsburg, Germany}

\author{Fulvio Parmigiani}
\affiliation{Dipartimento di Fisica, Università degli Studi di Trieste, 34127 Trieste, Italy}
\affiliation{Elettra-Sincrotrone Trieste S.C.p.A., 34149 Basovizza, Italy}
\affiliation{International Faculty, University of Cologne, Albertus-Magnus-Platz, 50923 Cologne, Germany}

\author{Massimo Capone} 
\email{capone@sissa.it}
\affiliation{International School for Advanced Studies 
	(SISSA) and CNR-IOM, Via Bonomea 265, I-34136, Trieste, Italy}

\author{Federico Cilento} 
\email{federico.cilento@elettra.eu}
\affiliation{Elettra-Sincrotrone Trieste S.C.p.A., 
34149 Basovizza, Italy}

\date{\today}

\begin{abstract}
FeSe$_{x}$Te$_{1-x}$ compounds present a complex phase diagram, ranging from the nematicity of FeSe to the $(\pi, \pi)$ magnetism of FeTe. We focus on FeSe$_{0.4}$Te$_{0.6}$, where the nematic ordering is absent at equilibrium. We use a time-resolved approach based on femtosecond light pulses to study the dynamics following photoexcitation in this system. The use of polarization-dependent time- and angle-resolved photoelectron spectroscopy allows us to reveal a photoinduced nematic metastable state, whose stabilization cannot be interpreted in terms of an effective photodoping. We argue that the 1.55 eV photon-energy-pump-pulse perturbs the $C_4$ symmetry of the system triggering the realization of the nematic state.
The possibility to induce nematicity using an ultra-short pulse sheds a new light on the driving force behind the nematic symmetry breaking in iron-based superconductors. Our results weaken the idea that a low-energy coupling with fluctuations is a necessary condition to stabilize the nematic order and ascribe the origin of the nematic order in iron-based superconductors to a clear tendency of those systems towards orbital differentiation due to strong electronic correlations induced by the Hund's coupling.
\end{abstract}

{\bf \maketitle }

%INTRO  %%%%%%%%%%%%%%%%%%%%%%%%%%%%%%%%%%%%%%%%%%%%%%%%%%%%%%%%%%%%%%%%%%%%%%%
\section*{INTRODUCTION}
Quantum-correlated materials based on transition metal atoms usually display rich phase diagrams where multiple electronic phases can be explored by tuning external parameters, like temperature or pressure. In the last years a novel handle to explore the complex properties of these materials has been found in time-resolved spectroscopies where the system is driven out of equilibrium by an ultrashort light pulse. In several cases, photoexcitation can trigger novel metastable transient states, whose existence could not be trivially inferred based on the equilibrium properties of the system.

In this perspective, time-resolved spectroscopies emerge as ideal tool to shed light on the competing orders characterizing the equilibrium phase-diagram of iron-based superconductors (IBS). In most members of the IBS family, by lowering the temperature, the undoped parent compounds show a structural-nematic transition from a tetragonal to an orthorhombic phase, often accompanied by a transition into a striped-antiferromagnetic metallic phase, while a superconducting phase emerges via doping or applying pressure \cite{Stewart_2011,Johnston_2010}. The interplay between superconductivity and the other instabilities is still debated.

In this work we study the electron dynamics of the FeSe$_{0.4}$Te$_{0.6}$ IBS using Time-Resolved ARPES (TR-ARPES). This technique, taking advantage of the selection rules of the photoemission process and the experimental geometry \cite{Damascelli_2003}, allows us to analyze the electron dynamics with orbital sensitivity, by performing polarization-dependent measurements. Our main result is the observation of a persistent modification of the electronic band structure, that indicates the realization of a new metastable condition triggered by the pump pulse. We argue that the new metastable state is the realization of a nematic phase in FeSe$_{0.4}$Te$_{0.6}$, in line with previous experimental indications of instability of the $C_4$ symmetry in this compound \cite{Singh_SciAdv2015}. We propose that electronic correlations induced by the Hund's coupling, which has been proved to strengthen the nematic orbital differentiation in the nematic phase of FeSe \cite{Fanfarillo_PRB2017}, stabilize the small anisotropy induced 
by the ultrafast excitation. 

In the last years FeSe$_{x}$Te$_{1-x}$ attracted a growing interest due to their anomalous phenomenology with respect to other members of the IBS family. The pure FeSe system undergoes a structural transition from the tetragonal to the orthorhombic phase around $T_s = 90$ K, but no long-range magnetic order develops upon lowering the temperature, while a superconducting phase appears below $T_c\sim 8$ K \cite{Coldea_Review2017}. Isovalent substitution of tellurium in place of selenium makes $T_c$ increase to $15$ K while suppressing the structural phase transition, which disappears around $x\sim 0.5$ \cite{Mizuguchi_Review2010}.

In this work we focus on samples with  $x=0.4$, for which  quasiparticle scattering microscopy clearly shows that the electronic structure remains unstable against a reduction of the symmetry from $C_4$ to $C_2$, i.e. with respect to a nematic symmetry breaking\cite{Singh_SciAdv2015}. Despite the absence of long-range  magnetic order, spin-fluctuations peaked at the nesting wave-vector between the hole and electron Fermi Surface (FS) are experimentally observed \cite{Qiu_PRL2009,Liu_NatMat2010,Christianson_PRB2013}, as well as short-range magnetic excitations at momentum $Q_m=(\pi,\pi)$ \cite{Bao_PRL2009, Katayama_JPSJ2010,Liu_NatMat2010, Lamura_JPCM2013}. Increasing further the Te-concentration suppresses the superconducting state around $x\sim 0.3$ and leads to the formation of a long-range magnetic order $Q_m=(\pi,\pi)$ characteristic of FeTe.

\begin{figure*}[tbh]
	\includegraphics[angle=0,width=1.\textwidth]{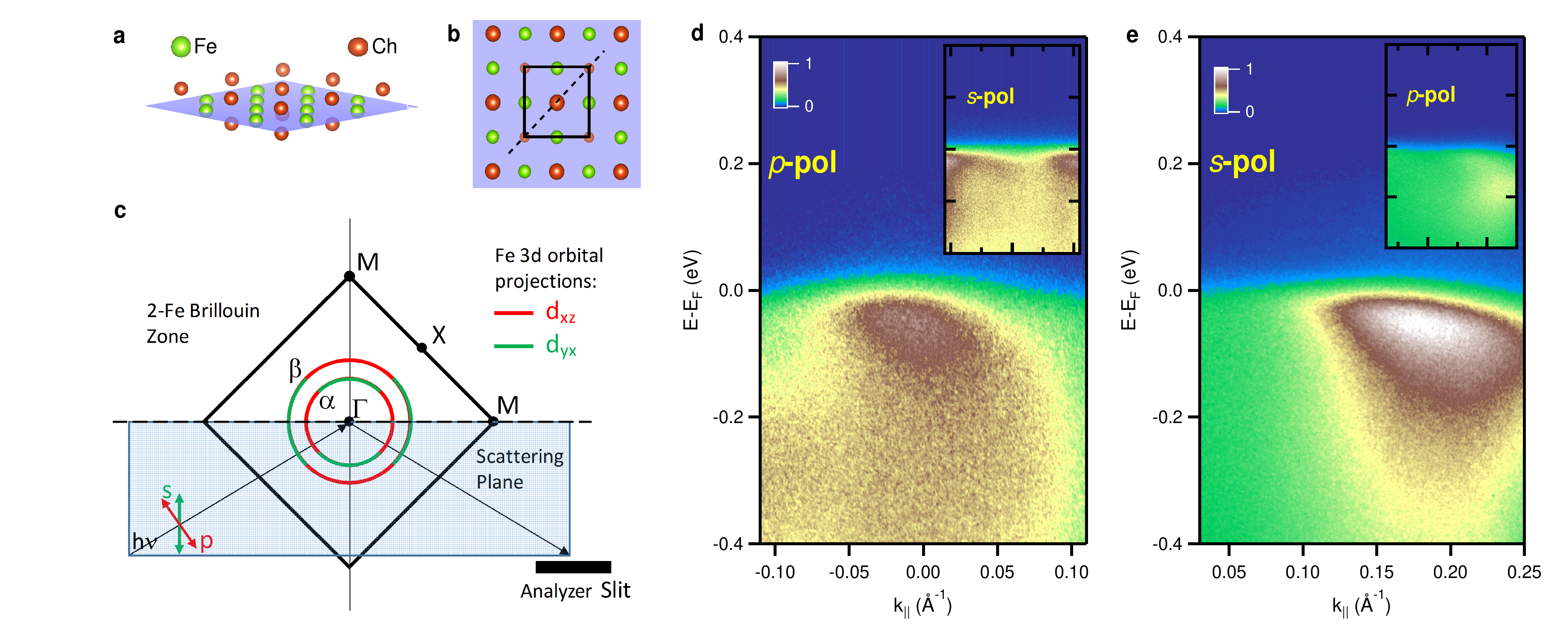} 
	\caption{FeSe$_{x}$Te$_{1-x}$ crystal structure, sketch of the experiment and polarization-dependent ARPES maps at equilibrium. {\bf a}, {\bf b} Crystal structure of a FeSe$_{x}$Te$_{1-x}$ layer and projection on the xy plane. The black square indicates the 2-Fe unit cell. {\bf c} Brillouin zone in the 2-Fe notation rotated by 45 degress with respect to panel b showing the geometry of the ARPES experiment. High-symmetry points are indicated. ARPES measurements have been performed along $\Gamma M$ (dashed black lines). In this experimental geometry
	%, owing to the band structure of FeSe$_{x}$Te$_{1-x}$, 
	the orbital character of the $\alpha$ ($\beta$) band is purely $d_{xz}$ ($d_{yz}$), as indicated by red (green) color. All the other bands are neglected for visualization clarity purpose. {\bf d} ARPES map acquired in $p$ polarization about $k=0$, evidencing the $\alpha$ band. In inset the ARPES intensity acquired in $s$ polarization is shown. {\bf e} ARPES map acquired in $s$ polarization centered about $k\sim0.16$ ${\AA}^{-1}$ evidencing the $\beta$ band. In inset the ARPES intensity acquired in $p$ polarization is shown. Insets share the energy, momentum and intensity scales of the corresponding main panel.} 
\label{EqARPES}
\end{figure*} 

From a theoretical point of view, the complex phenomenology observed in the FeSe$_{x}$Te$_{1-x}$ system by varying the Te-concentration appears as a serious obstacle for the existence of a universal scenario for the IBS. Within low-energy approaches, in fact, the ordered phases of IBS are originated by FS instabilities resulting from the exchange of low-energy collective modes (with either orbital or spin character) between hole and electron pockets connected by a nesting vector\cite{Chubukov_Chapter2015, Onari_Chapter2015}. However, while a low-energy approach can still explain the presence of a nematic phase without magnetism on the FeSe side \cite{Fanfarillo_PRB2018}, the appearance of a magnetic phase at a momentum different from the nesting one on the FeTe side can hardly be explained by any modelling based exclusively on the low-energy FS properties.

The discussion about the most appropriate low-energy description for FeSe-based systems cannot neglect the role of electron-electron interactions, which are hardly negligible in IBS and are particularly strong in chalcogenides, as theoretically discussed, e.g., in \cite{Aichhorn_old,Lanata_PRB2013} and experimentally found in \cite{Arcon_PRB2010, Xu_PRB2011, Tamai_PRL2010, Ieki_PRB2014, Yi_NatComm2015, Ambolode_PRB2015}. Analogously to other IBS, correlations in FeSe-based materials are mainly arising from local electronic interactions, and in particular by the Hund's coupling \cite{DeMedici_PRL2014}. A number of experiments \cite{Wang_2014, Ding_2014, Li_2014, Gao_2014} place FeSe-based systems close to a crossover between a standard metal and a Hund's metal, where theoretically we expect that the suppression of the overall coherence of conduction electrons, is accompanied by a simultaneous effect of orbital differentiation \cite{DeMedici_PRL2014} which descends from an effective decoupling between the orbitals \cite{DeMedici_PRL2014,Fanfarillo_PRB2015}.

The theoretical study of strong electronic correlations in IBS has been so far widely devoted to the properties of the metallic phase and the relationship between the emergence of orbital-selective correlations and the low-temperature instabilities has been touched upon only recently. In particular, in Ref. \cite{Fanfarillo_PRB2017}, the analysis of the nematic susceptibility in the correlated regime reveals that, within the correlated description, the nematic order is not spontaneously established, i.e. we need to explicitly break the $C_4$ symmetry to trigger it. However, once the symmetry is broken, electronic correlations induced by Hund's coupling cooperate to make the nematic order stable and enhance the differentiation between the  $xz$ and $yz$ orbitals, thereby increasing the nematic character.
%RESULTS  %%%%%%%%%%%%%%%%%%%%%%%%%%%%%%%%%%%%%%%%%%%%%%%%%%%%%%%%%%%%%%%%%%%%%
\section*{RESULTS}
\begin{figure*}[tbh]
\includegraphics[angle=0,width=0.93\textwidth]{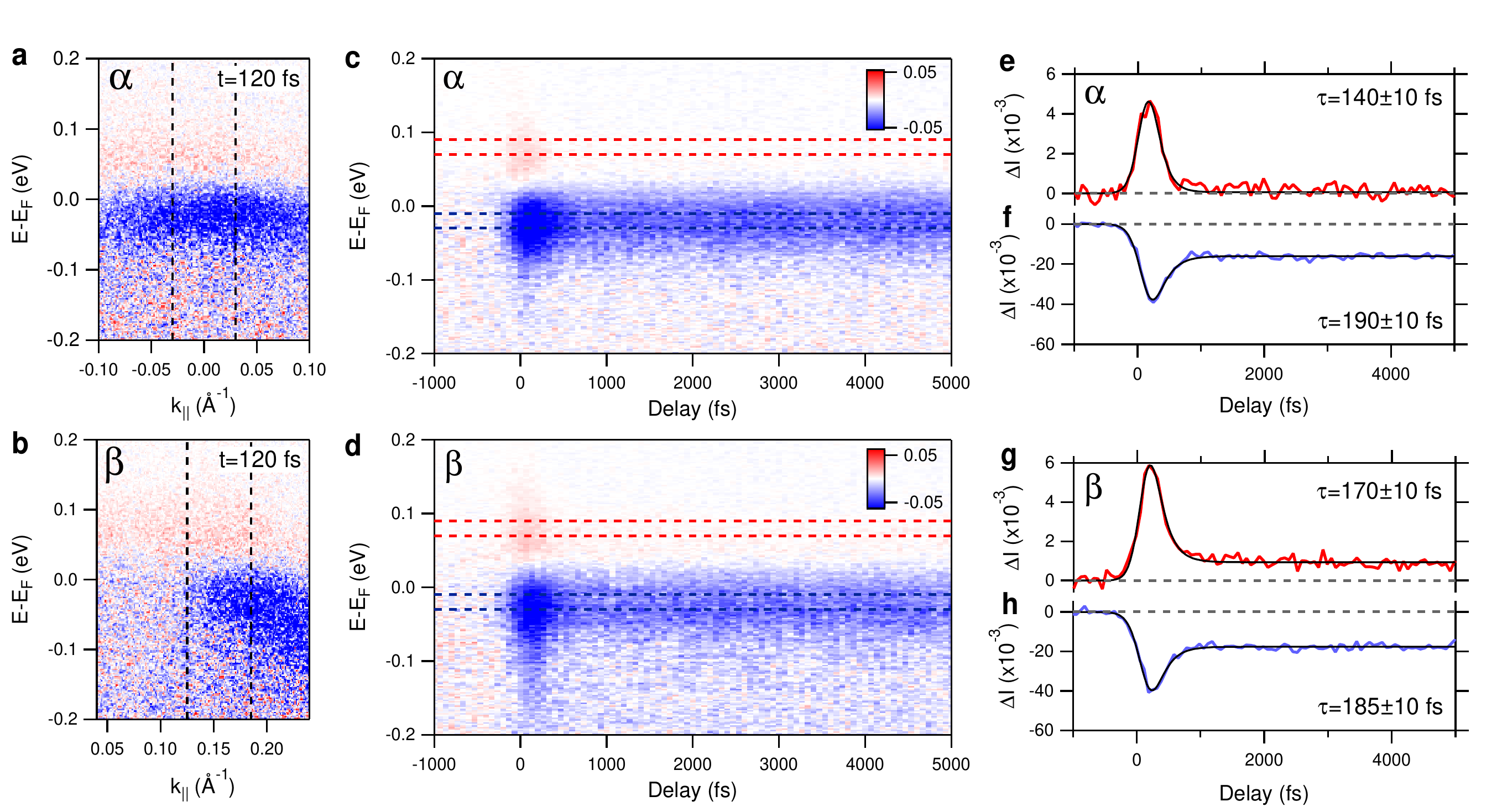} 
\caption{Out-of-equilibrium ARPES maps and electron dynamics. {\bf a}, {\bf b} Differential ARPES maps for the $\alpha$ and $\beta$ bands, collected 120 fs after excitation. The dashed black lines indicate the momentum region in which the EDCs shown in panels {\bf c}, {\bf d} for all pump-probe delays have been extracted. Panels a-d share the same colorscale. {\bf e}, {\bf f} ($\alpha$ band) and {\bf g}, {\bf h} ($\beta$ band) show respectively the intensity integrated in a 20 meV wide energy window centered either at +80 meV above $E_F$ (indicated by red dashed lines in panels c,d or at -20 meV below $E_F$ (indicated by blue dashed lines in panels c,d). Solid black lines in panels e-h are the fit to the data (see main text). The time-constants extracted are reported on each graph. Both bands display a persistent signal below $E_F$. Above the Fermi energy, only the $\beta$ band present the same persistent signal.}
\label{IntDynamics}
\end{figure*}

In order to set the notations, we report in Fig.~1a the basic crystal structure of a FeSe$_{x}$Te$_{1-x}$ layer, with the chalcogen atoms alternating above and below the Fe-atoms plane. Fig.~1b shows the projection on the $xy$ plane with the indication of the 2-Fe unit cell (black square). 
%This defines the 2-Fe Brillouin zone, which is shown in Fig.~1c (rotated by 45 degrees). 
Fig.~1c represents a sketch of the experimental geometry describing the sample alignment with respect to the electron analyzer slit: equilibrium ARPES measurements have been performed along the Fe-Fe direction (dashed line of panels b,c), corresponding to the high-symmetry direction $\Gamma M$ in the 2-Fe unit cell notation. At $h\nu=6.2$ eV probe photon energy, only the momentum-space region near the $\Gamma$ point can be accessed, where the hole-like bands are located. The two innermost bands, named $\alpha$ and $\beta$ (not to scale in Fig.~1c), display a single $3d$ orbital component along the $\Gamma M$ high-symmetry direction \cite{Moreschini_PRL2014}, which can be selectively probed thanks to the polarization-dependence of the ARPES matrix elements (see TABLE I).

The polarization-dependent ARPES maps collected on FeSe$_{0.4}$Te$_{0.6}$ are shown in Fig. 1d, 1e, for $p$ and $s$ polarization, respectively. We clearly resolve and isolate two hole-like bands, $\alpha$ and $\beta$, lying close to the Fermi level. The $\alpha$ band, revealed in $p$ polarization, is touching the Fermi energy, $E_F$, and is centered at $k=0$; the $\beta$ band, revealed in $s$ polarization, crosses the Fermi level at $k_F \sim 0.12$ ${\AA}^{-1}$. We do not observe the third hole-like band $\gamma$, predicted by band calculations \cite{Chen_PRB2010, Moreschini_PRL2014} and experimentally detected in \cite{Chen_PRB2010, Tamai_PRL2010, Nakayama_PRL2010, Miao_PRB2012, Ieki_PRB2014, Yi_NatComm2015, Ambolode_PRB2015}, since it crosses the Fermi level at a larger wave-vector, beyond the region of momenta we can access. 
The direct comparison between the main panels of Fig. 1d and 1e and their insets, where we use the opposite polarization with respect to the main panel in the same energy and momentum ranges, clearly proves the capability to selectively probe the $\alpha$ or $\beta$ bands by proper choices of the probe polarization state. In 
fact, the $\alpha$ band signal is completely suppressed in $s$ polarization (inset of Fig. 1d), while we recorded only a weak intensity in $p$ polarization (inset of Fig. 1e) at $k\sim0.2$ ${\AA}^{-1}$ with the $\beta$ band signal around $k_F$ completely suppressed. From this picture we learn that the $\alpha$ ($\beta$) band has even (odd) initial state with respect to the scattering plane, in agreement with previous polarized ARPES studies \cite{Chen_PRB2010, Tamai_PRL2010, Miao_PRB2012}. The result is consistent with band calculations giving the $\alpha$ and $\beta$ bands defined at low-energy from the even and odd combination of $xz/yz$ orbitals \cite{Chen_PRB2010, Moreschini_PRL2014}, with the $\alpha$ ($\beta$) band purely $xz$ ($yz$) along $\Gamma-M$ (see TABLE I). 
\begin{table}[t]
\begin{center}
\begin{tabular}{cccccccc} 
{High-Symmetry}&\ {Experimental} \  & \multicolumn{5}{c}{ $3d$-orbitals}\\
direction & geometry & $d_{z^2}$ & $d_{x^2-y^2}$ &  $d_{xz}$ \ & $d_{yz}$ \ &$d_{xy}$ \\ 
\hline 
$\Gamma-M$ & $p$ &  $\checkmark$ & $\checkmark$ & $\checkmark$ & & \\ 
&$s$ & & & & $\checkmark$& $\checkmark$ \\  
\hline 
\end{tabular} 
\caption{Polarization-resolved ARPES, using $p$ (horizontal) or $s$ (vertical) polarization, allows to disentangle the Fe $3d$-orbitals contribution when measuring along the $\Gamma M$ high symmetry direction.} 
\end{center}
\vspace{-0.5cm}
\end{table}
The residual spectral weight appearing below $E_F$ in $p$ polarization around $k\sim 0.2$ ${\AA}^{-1}$ (inset of Fig. 1e) is compatible with the presence of $z^2$ orbital states below the Fermi level, as also discussed in \cite{Tamai_PRL2010, Ieki_PRB2014} and in agreement with the selection rules summarized in TABLE I. 

Having clarified the nature of the electronic bands measured by ARPES and their orbital character, we turn to analyze the non-equilibrium dynamics. TR-ARPES measurements are performed in the same geometry used for the experiments at equilibrium. The photoexcitation at $h\nu_{Pump}=1.55$ eV (in $p$ polarization) is collinear to the probe pulse, and the absorbed fluence is set to $\sim$200$\pm$25 $\mu J/cm^2$. The non-equilibrium signal is uniquely related to a single orbital according the selection rules of the photoemission process discussed within the equilibrium measurements. Fig. 2 displays the non-equilibrium electron dynamics for the $\alpha$ and $\beta$ bands. 
\begin{figure*}[tbh]
\includegraphics[angle=0,width=0.88\textwidth]{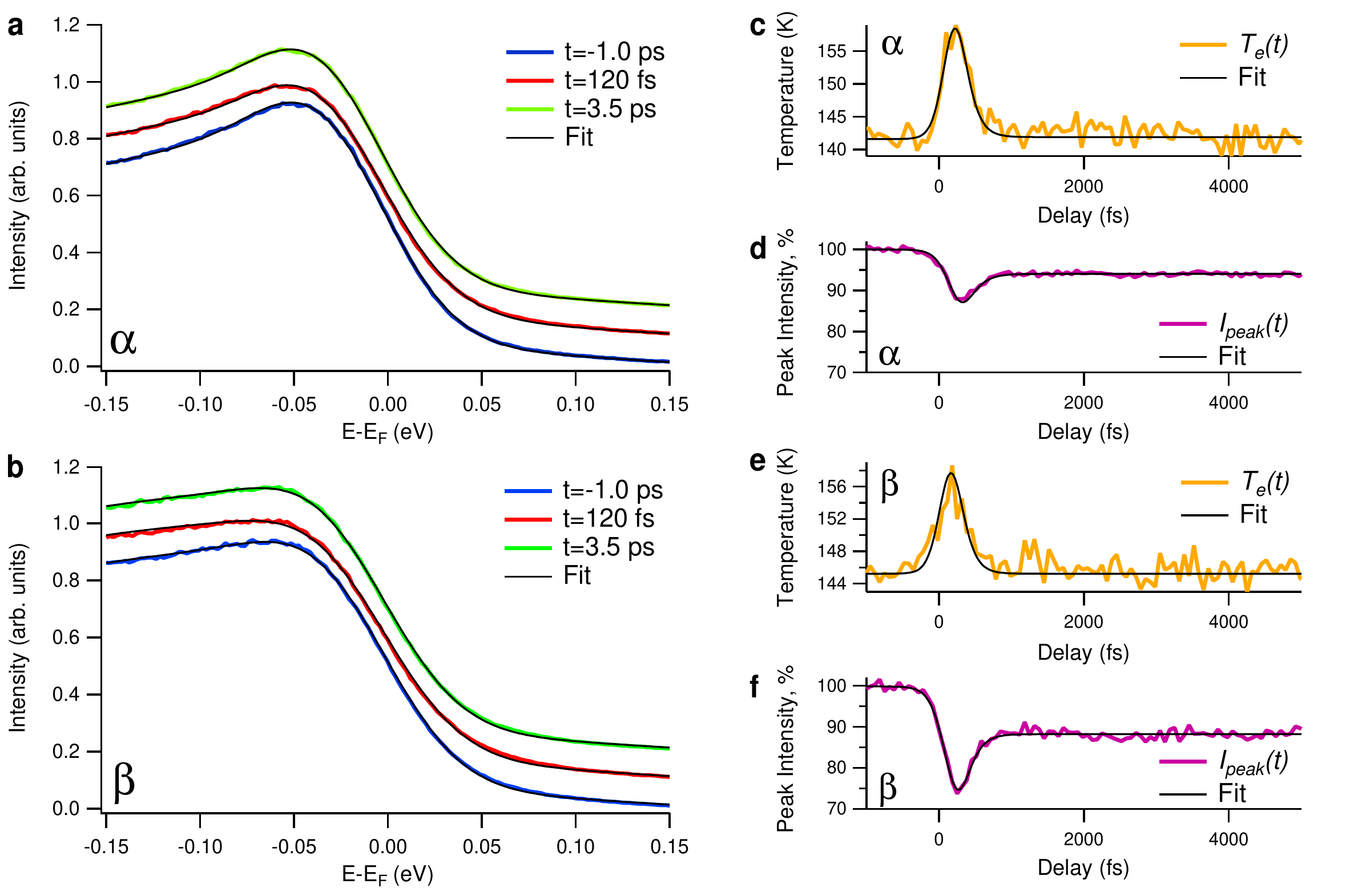} 
\caption{Lineshape analysis. {\bf a}, {\bf b} Energy Distribution Curves collected at pump probe delays $t=-1$ ps (before the pump), $t=120$ fs (just after $t_0$) and $t=3.5$ ps (after the pump) for the $\alpha$ and $\beta$ bands respectively. %Black lines are the fit to the data (see main text). 
{\bf c}, {\bf d} $T_e(t)$ (orange) and $I_{peak}(t)$ (violet) for the $\alpha$ band. {\bf e}, {\bf f} $T_e(t)$ (orange) and $I_{peak}(t)$ (violet) for the $\beta$ band. %In all panels, black lines are the fit to the data. 
For both $\alpha$ and $\beta$, $T_e(t)$ displays an impulsive behavior and relaxes to the initial value. Conversely, $I_{peak}(t)$, after a fast depletion, saturates to $\sim94\%$ and $\sim88\%$ of the initial value for the $\alpha$ and $\beta$ bands respectively.}
\label{IntEDCs}
\end{figure*}

Fig.~2a, 2b show the differential ARPES maps for the $\alpha$ and $\beta$ bands respectively, collected $120$ fs after the excitation. The maps has been obtained by subtracting from the map acquired at pump-probe delay $t=120$ fs the corresponding map acquired before the arrival of the pump pulse (at $t=-1$ ps). A reduction of the photoemission intensity is depicted in blue, and indicates the depletion of the occupied states, while the red color indicates an increase of intensity with respect to equilibrium. Fig.~2c, 2d show the time-dependent $differential$ Energy Distribution Curves (EDCs) averaged over the momenta falling within the range ($0.06$ ${\AA}^{-1}$ wide) indicated with black dashed lines in 2a-b, for the $\alpha$ and $\beta$ bands respectively. The differential EDCs clearly present two phenomena. The first, occurring just after the excitation, consists in an impulsive signal decaying on an ultrafast timescale. The second develops for $t\gtrsim500$ fs and mainly affects the states below $E_F$. The first effect can be attributed to a population effect and can be mimicked by an instantaneous increase of the electronic temperature, while the second consists in a persistent modification of the photoemission intensity, which lasts for several tens of picoseconds, and indicates a metastable modification of the electronic properties. The electron dynamics extracted above and below $E_F$ in the intervals $80\pm10$ meV (in red) and $-20\pm10$ meV (in blue) are shown in Fig.~2e-f and 2g-h for the $\alpha$ and $\beta$ band respectively. The fitting analysis of the traces, performed using a single-exponential decay convoluted with the experimental time-resolution of $200$ fs (mimicked by a Gaussian), confirms the qualitative picture. At first, we observe a fast decay of both the time-resolved signal above and below $E_F$. Then, we find a large persistent signal for both bands below $E_F$, and for the $\beta$ band also above $E_F$.

To quantify those effects and disentangle the impulsive temperature modification from the persistent modification of the spectral weight, we carefully analyze the lineshape of EDCs curves as a function of pump-probe delay. For this reason, here we average over a tiny $0.01$ ${\AA}^{-1}$ momentum range. This choice makes the weak signal measured above $E_F$ at large pump-probe delays not detectable. As a consequence, this signal will not enter in the analysis described below. Fig.~3a, 3b show, for the $\alpha$ and $\beta$ bands respectively, three representative EDCs collected at pump probe delays $t=-1$ ps, $120$ fs, $3.5$ ps. We model the photoemission intensity along each EDC at fixed pump-probe delay as:
\begin{equation}\label{Photoem}
f(E)= \frac{1}{e^{(E-E_{F})/k_B T_e}+1} \cdot (L_{peak} +L_{tail}) 
\end{equation}
where $k_B$ is the Boltzmann constant and $T_e$ is the effective electronic temperature. $L_i$ are Lorentzian functions. To reproduce the EDCs we use two Lorentzian functions: $L_{peak}$, for the lineshape, and $L_{tail}$ for the high-binding-energy tail. The parameters of $L_{tail}$ are obtained from the fit of the EDC at the equilibrium, and are kept fixed for the subsequent analysis. For all the pump-probe delays the lineshape evolution can be reproduced by changing the intensity, $I_{peak}$, of $L_{peak}$ only. The description of the electronic population through the Fermi-Dirac function is satisfactory at all pump-probe delays by invoking a modification of the electronic temperature $T_e$ alone, with no change of the chemical potential. Hence, the time-dependent EDCs for both the $\alpha$ and $\beta$ bands can be reproduced at all delays by only two parameters, namely, $I_{peak}(t)$ and $T_e(t)$. Anyhow, because of the broad spectral features and the energy resolution of the TR-ARPES experiment, we cannot exclude that the modification of the intensity $I_{peak}(t)$ arises from a shift (or splitting) of the bands that affects the Fermi wavevector, which represents a fingerprint of the nematic order in FeSe \cite{Coldea_Review2017}.

In Fig.~3a, 3b the fits (solid black lines) are superimposed to the curves, 
showing a good overall agreement. $T_e(t)$ (orange curves) and $I_{peak}(t)$ (violet curves) obtained from the analysis of all the pump-probe delays are shown in Fig.~3c, 3d and Fig.~3e, 3f for the $\alpha$ and $\beta$ band respectively. 
$T_e(t<t_0)$ is slightly larger than the target sample temperature (T$\sim130$ K), for both bands. This effect can be ascribed to the heating of the sample induced by the pump pulse power, of the order $\sim10$ mW. The small difference ($<4$ K) between the $T_e(t<t_0)$ of the two band can be related to both temperature fluctuations in the experiments and the uncertainty in the analysis, due to the small non-equilibrium signal. For both bands, the time evolution of the effective temperature, $T_e(t)$, displays an impulsive behavior, with an increase of the order 10-15 K, followed by a fast relaxation to values close to the initial one. 
The main result of this work comes from the analysis of the time evolution of $I_{peak}(t)$, shown in Fig.~3d, 3f. This quantity is expressed as a percentage of the intensity of the peak at time $t$ with respect to the value determined for $t=-1$ ps. We first observe a reduction of the 
intensity characterized by a fast dip, having timescale of the same order of the experimental time resolution. The dip is followed by a steady condition that 
lasts for several tens of picoseconds for both $\alpha$ and $\beta$, confirming the qualitative observation of a metastable state shown in Fig. 2. Notice that, after $t\approx1$ ps from the excitation, the intensities of the peaks for the two bands saturate to different values with respect to their equilibrium intensity, namely $\sim94\%$ and $\sim88\%$ for the $\alpha$ and $\beta$ bands respectively. This tiny difference is a robust feature of the system dynamics that we observe in every measurements and reveals that the persistent signal observed below $E_F$ presents a differentiation in the populations of the $\alpha$ and $\beta$ bands. It is worth noticing that even if it was not possible to perform a quantitative analysis of the persistent signal that we appreciate above $E_F$ in Fig.~2g, we find that it goes in the same direction of the differentiation found below $E_F$ in terms of orbital populations. In fact, the $\beta$ band that below $E_F$ recovers only the $\sim88\%$ of its equilibrium population, is the one that consistently shows the strongest 
persistent signal above $E_F$ (see Fig.~ 2g vs 2e).

%DISCUSSION %%%%%%%%%%%%%%%%%%%%%%%%%%%%%%%%%%%%%%%%%%%%%%%%%%%%%%%%%%%%%%%%%%%
\section*{DISCUSSION}

The analysis of the TR-ARPES data clearly shows that an ultrafast excitation can trigger a metastable state which affects mainly the electronic states below the Fermi level and, most importantly, survives on a picosecond timescale.
It is worth mentioning that at the temperature of our experiment ($130$ K) and at higher temperatures, FeSe$_{0.4}$Te$_{0.6}$ presents a tetragonal structure and it is a standard paramagnet characterized by a bad metallic behavior as other members of the IBS family. Even by lowering the temperature this compound does not display any phase transition i.e. it does not become magnetic as FeTe neither nematic as FeSe. Thus, the photoinduced metastable state appears to induce in FeSe$_{0.4}$Te$_{0.6}$ a new phase not present at equilibrium.

The analysis of the persistent signal gives indications about the nature of the metastable phase. We found, in fact, a tiny, but robust, differentiation in the populations of the $\alpha$ and $\beta$ bands. Since along $\Gamma M$ the $\alpha$ and $\beta$ bands are composed by $xz$, and $yz$ orbitals respectively, this means that the metastable phase presents a differentiation in the orbital populations, similarly to what observed in the nematic phase of FeSe \cite{Huh_arxiv2019}. 
Notice that the orthorhombic phase transition found in pure FeSe is suppressed upon Te-doping and completely disappears around $x=0.5$ in FeSe$_x$Te$_{1-x}$ \cite{Mizuguchi_Review2010}. There are however evidences that FeSe$_{0.4}$Te$_{0.6}$ strongly responds to a reduction of symmetry from $C_4$ to $C_2$, as shown in Ref. \cite{Singh_SciAdv2015}, where quasiparticle scattering measurements showing $x/y$ anisotropic signals at low temperature have been interpreted as the evidence of electron nematicity probably triggered by strain induced upon cooling.

The small unbalance in the orbital population of the nematic metastable phase observed in our compound is consistent with ARPES observations in FeSe \cite{Coldea_Review2017} 
that  found  $xz$/$yz$ orbital splitting at the symmetry points for both hole and electron bands, with a sign change of the orbital polarization in momentum space \cite{Suzuki_PRB2015, 
Zhang_PRB2016, Fanfarillo_PRB2016, Yi_arxiv2019, Huh_arxiv2019}. Small values of the imbalance between the population of $xz$/$yz$ orbitals are also theoretically expected in systems, as IBS, characterized by a sizeable value of the Hund's coupling, that favors configurations where the orbital occupation is uniform in order to locally maximize the total spin. 
Among different nematic orderings, in fact, only those with small charge unbalance between the $xz$/$yz$ orbitals can survive in a correlated regime dominated by $J_H$, as shown in \cite{Fanfarillo_PRB2017,Qimiaofurbo}. For this kind of orders, the effect of a generic small $xz$/$yz$ nematic perturbation, is found to be enhanced by correlations induced by $J_H$. These results highlights a non-trivial interplay between the trigger which breaks the $C_4$ symmetry and Hund's driven correlations which enhance the differentiation between the orbitals and consequently the nematic character of the system. We propose that this mechanism is behind the photoinduced nematic state found in our time-resolved experiments and that, while in equilibrium the breaking of the symmetry is naturally associated with spin and/or orbital fluctuations, in the present non-equilibrium results it can be caused by the pulse driving the system out of equilibrium. 
An ultra-short pulse can in fact explicitly break the $C_4$ symmetry randomly exciting the $xz/yz$ electronic orbital states and, once the symmetry is broken, electronic correlations can cooperate to enhance such differentiation and make this phase metastable. 

In the following we want to corroborate this interpretation quantitatively. The ultra-short pulse cannot excite selectively any orbital or any particular symmetry, thus it can be modelled as a generic linear combination of any possible perturbation of the electronic structure of the system as:
\be
P = \sum_i  a_i \ P^i_{C_4} +  b_i  \ P^i_{C_2} 
\ee
where $a_i$ and $b_i$ are the coefficients of the linear combination and $ P^i$ the perturbations. We distinguish between perturbations preserving the $C_4$ symmetry $P^i_{C_4}$ and nematic perturbations as e.g. $ P^i_{C_2}   \sim  \Delta n_{\bf k} = n_{xz}({\bf k}) - n_{yz}({\bf k})$, characterized by an occupation unbalance between the $xz/yz$ orbitals. 
The analysis of \cite{Fanfarillo_PRB2017} demonstrates that, among all the nematic perturbations, only perturbations with a sign modulation in the momentum space survive in the correlated regime. For all these perturbations, that we indicate as Sign-Change Orbital (SCO) perturbations $P^{SCO}_{C_2}$, the unbalance $\Delta n$ introduced by the SCO component is maintained almost unchanged by interactions, while a strong enhancement of the orbital differentiation is found.
This effect can be quantified by considering the orbital quasiparticle spectral weight $Z_{orb}$, that is equal to one in non-interacting systems and decreases with increasing correlations. 
In the $C_4$ symmetric phase the $xz/yz$ orbital are equivalent, i.e. $Z_{xz}=Z_{yz}$. The nematic perturbation $P^i_{C_2}$ introducing a finite population unbalance $\Delta n_{\bf k}$ induces a differentiation $\Delta Z = Z_{xz} - Z_{yz}$. By studying the dependence of $\Delta Z$ from local interactions, it has been shown \cite{Fanfarillo_PRB2017} that $\Delta Z$ grows monotonously with the interactions and presents a maximum for interactions values corresponding to the crossover between the weakly correlated metal and the strongly correlated regime. We can estimate the role of the Hund's coupling in inducing such enhancement by computing the orbital differentiation in the correlation strength $\Delta Z$ for different values of the Hund's coupling $J_H$. Using the numerical results of \cite{Fanfarillo_PRB2017} we find that $\Delta Z_{J_H \neq 0}/ \Delta Z_{J_H=0}$ is always larger than 1 and at the crossover it is around 70. This result indicates that the Hund's induced correlations makes the system extremely sensitive towards an orbital nematic differentiation. We argue that, once the ultra-short pulse perturbs the system inducing a small difference in the population of the $xz/yz$ orbitals, the correlations induced by local interactions stabilize that phase maintaining unchanged the small population unbalance $\Delta n $, and enhancing the correlation strength orbital differentiation $\Delta Z $. 

The above picture is consistent with the experimental result discussed in this work that highlights the creation via ultra-short pulse of a small $\a/\b$ population unbalance that persist on a timescale of the order of picoseconds. The analysis of the dynamics displayed by the electron pockets at $M$ would give more information about the nature of the metastable nematic state, including its symmetry. This would provide a verification of the cooperative mechanism we proposed since, within a correlated scenario, a SCO polarization creating a small charge imbalance in the orbital population is the only perturbation expected to survive and be stable in a correlated system as FeSe$_x$Te$_{1-x}$ \cite{Fanfarillo_PRB2017}.

It is finally worth to notice that any chemical doping starting from stoichiometric compounds of the IBS family results in a reduction of the nematic ordering temperature, in striking contrast with the light-induced nematic transient state reported here. This clearly rules out an interpretation of the present experiment in terms of a photodoping of the FeSe$_{0.4}$Te$_{0.6}$ band structure.
%CONCLUSIONS %%%%%%%%%%%%%%%%%%%%%%%%%%%%%%%%%%%%%%%%%%%%%%%%%%%%%%%%%%%%%%%%%%%
\section*{CONCLUSIONS}
We performed an orbital-resolved study of the electron dynamics of FeSe$_{0.4}$Te$_{0.6}$ by means of TR-ARPES. At equilibrium, we resolve two hole-like bands composed by $xz$ and $yz$ orbitals crossing the Fermi level close to the $\Gamma$-point. Time-resolved measurements clearly identify for both bands two separate effects: the instantaneous modification of the electronic temperature and the modification of the photoemission intensity of the bands, mainly affecting the electronic states below the Fermi energy. The latter effect persists on a picosecond timescale and indicates that the ultra-short pulse triggers the creation of a metastable state.
The quantitative analysis of the persistent signal reveals the nematic character of such a phase. The signal is in fact characterized by a population unbalance between the $\a/\b$ bands, i.e. between the $xz/yz$ orbital states composing those bands along the $\Gamma M$ high-symmetry direction measured in the experiment. The stabilization of such a nematic state is clearly in contrast with an effective photodoping, which is expected to reduce the nematic ordering in analogy with chemical doping.

The possibility of inducing a metastable phase having nematic character in FeSe$_{0.4}$Te$_{0.6}$ using an ultra-short pulse can help to shed light on the driving force behind the nematic symmetry breaking in IBS. Our results are compatible with a scenario in which spin fluctuations peaked at the nesting vector mediate nematic ordering in undoped FeSe, while Te-doping weakens these fluctuations in favor of $(\pi,\pi)$ ordering and as a consequence nematicity disappears. In our sample, which lies close to this crossover, nematicity turns out to be restored by an impulsive excitation which breaks the rotational symmetry. In that respect, our results weaken the idea that low-energy coupling with fluctuations is a necessary condition to stabilize the nematic order.
It is indeed unlikely that a pulsed photoexcitation could trigger a collective (orbital or spin) mode capable to induce the nematic order, while it is more plausible that the pulse randomly perturbs the electronic structure of the system breaking accidentally the $C_4$ symmetry. We argue that, once the pulse breaks the symmetry, electronic correlations induced by Hund's coupling cooperate to stabilize the metastable conditions triggered by the pulse. 

FeSe$_{0.4}$Te$_{0.6}$ \cite{Singh_SciAdv2015} and more in general IBS \cite{Pal} appear to be systems prone to a nematic symmetry breaking. Our study ascribes the origin of such effect to a clear tendency towards orbital differentiation due to electronic correlations induced by the Hund's coupling. In this scenario, strain, spin- or orbital-fluctuation anisotropy or ultra-short pulse excitation represent just the trigger to realize such a tendency.\\

Note added: In the last phase of preparation of the article we became aware of the results of a recent pump and probe experiment analyzing the dynamics of the nematic orbital order in FeSe \cite{Shimojima_NatComm2019}, where the femtosecond optical pulse is used to perturb the electronic nematic order of FeSe and study the ultrafast dynamics of electronic nematicity, finding a rapid destruction of nematic ordering. This is not in contrast with the experimental results reported in this work where starting from the paramagnetic FeSe$_{0.4}$Te$_{0.6}$ we observe the appearance of a persistent nematic state and it is fully compatible with our theoretical picture, which does not predict a pump-induced enhancement of a pre-existing nematic order.

%SYSTEM AND EXPERIMENT DETAILS %%%%%%%%%%%%%%%%%%%%%%%%%%%%%%%%%%%%%%%%%%%%%%%%
\section*{METHODS}
High quality single crystals of FeSe$_{0.4}$Te$_{0.6}$ were grown by the 
self-flux method, as discussed in \cite{Tsurkan_2011}. ARPES and TR-ARPES measurements were performed at the T-ReX Laboratory (Elettra, Trieste), 
using $h\nu=6.2$ eV as a probe photon energy. This has been obtained as the 
fourth harmonic of the fundamental laser output of an ultrafast Ti:Sapphire 
regenerative amplifier (Coherent RegA), producing $\sim50$\,fs pulses at 
$\lambda\sim800$\,nm directly used as a pump. The photoelectrons are detected by a Specs Phoibos 225 hemispherical electron analyzer. The energy and momentum resolutions of the setup are respectively $\sim50$ $meV$ and $\sim0.005$ ${\AA}^{-1}$. The sample is mounted on a six-degrees of freedom cryomanipulator. The polarization state of the probe beam can be set to either $p$ state (horizontal) or $s$ state (vertical), using a $\lambda/2$ waveplate. The slit of the electron analyzer is horizontal. The temperature throughout the experiments was set to $\sim130$ K. 

\section*{ACKNOWLEDGEMENTS}
L.F. acknowledges E. Bascones and B. Valenzuela for useful discussions and constructive comments. M.C. acknowledges  financial  support from  MIUR  PRIN 2015 (Prot.2015C5SEJJ001) and SISSA/CNR project ``Superconductivity, Ferroelectricity and Magnetism in bad metals'' (Prot.  232/2015).

\section*{COMPETING INTERESTS}
The authors declare no competing interests.

\section*{AUTHOR CONTRIBUTIONS}
F.C., D.K., A.S., G.M., and A.C. performed TR-ARPES experiments. F.C. and D.K. analyzed the data. L.F. and M.C. elaborated the theoretical support. V.T., D.C. and J.D. grew and characterized samples. F.C., L.F., D.K., F.P. and M.C. wrote the manuscript. All authors discussed and contributed to the manuscript. F.C. and F.P. designed the experiment.

\section*{DATA AVAILABILITY}
The data supporting the findings of this study are available from the corresponding authors upon reasonable request.

\bibliography{FeSeTe}

\end{document}